\documentclass[%
reprint,
superscriptaddress,
showpacs,preprintnumbers,
 amsmath,amssymb,
 aps,
 longbibliography,
prb,
floatfix,
]{revtex4-1}

\usepackage{epsfig}
\usepackage{subfigure}
\usepackage{graphicx}
\usepackage{dcolumn}
\usepackage{bm}
\usepackage{float}
\usepackage{color}
\usepackage[normalem]{ulem}
\usepackage{soul}
\usepackage{hyperref}
\hypersetup{
    unicode=false,          
    pdftoolbar=true,        
    pdfmenubar=true,        
    pdffitwindow=false,     
    pdfstartview={FitH},    
    pdftitle={My title},    
    pdfauthor={Author},     
    pdfsubject={Subject},   
    pdfcreator={Creator},   
    pdfproducer={Producer}, 
    pdfkeywords={keyword1} {key2} {key3}, 
    pdfnewwindow=true,      
    colorlinks=true,       
    linkcolor=blue,          
    citecolor=blue,        
    filecolor=blue,      
    urlcolor=blue,       
    plainpages=false        
}
\usepackage{comment}
\usepackage{xcolor}
\usepackage{physics}

\begin{document}

\title{Frustrated Magnetism from Local Moments in FeSe}
\author{Harrison Ruiz}
\affiliation{Department of Physics, Stanford University, Stanford, California 94305, USA}
\affiliation{Stanford Institute for Materials and Energy Sciences, SLAC National Accelerator Laboratory and Stanford University, Menlo Park, CA 94025, USA}
\author{Yao Wang }
\affiliation{Stanford Institute for Materials and Energy Sciences, SLAC National Accelerator Laboratory and Stanford University, Menlo Park, CA 94025, USA}
\affiliation{Department of Applied Physics, Stanford University, Stanford, California 94305, USA}
 \affiliation{Department of Physics, Harvard University, Cambridge 02138, USA}
\author{Brian Moritz}
\affiliation{Stanford Institute for Materials and Energy Sciences, SLAC National Accelerator Laboratory and Stanford University, Menlo Park, CA 94025, USA}
\affiliation{Department of Physics and Astrophysics, University of North Dakota, Grand Forks, North Dakota 58202, USA}

\author{Andreas Baum}
\affiliation{Walther Meissner Institut, Bayerische Akademie der Wissenschaften, 85748 Garching, Germany}
\affiliation{Fakult\"{a}t f\"{u}r Physik E23, Technische Universit\"{a}t M\"{u}nchen, 85748 Garching, Germany}
\author{Rudi Hackl}
\affiliation{Walther Meissner Institut, Bayerische Akademie der Wissenschaften, 85748 Garching, Germany}

\author{Thomas P. Devereaux}
\affiliation{Stanford Institute for Materials and Energy Sciences, SLAC National Accelerator Laboratory and Stanford University, Menlo Park, CA 94025, USA}

\date{\today}
\begin{abstract}
  We investigate properties of a spin-1 Heisenberg model with extended and biquadratic interactions, which captures crucial aspects of the low energy physics in FeSe. While we show that the model exhibits a rich phase diagram with four different magnetic ordering tendencies, we identify a parameter regime with strong competition between N\'{e}el, staggered dimer, and stripe-like magnetic fluctuations, accounting for the physical properties of FeSe. We evaluate the spin and Raman responses using exact diagonalization. Through comparison with experiments we find enhanced magnetic frustration between N\'{e}el and co-linear stripe ordering tendencies, which increases with increasing temperature. The explanation of these spectral behaviors with this frustrated spin model supports the idea of local spin interactions in FeSe.\end{abstract}

\maketitle

\section{Introduction}

Magnetic excitations are believed to play a significant role in the high-$T_c$ copper and iron-based superconductors.\cite{Scalapino:2012,Li:2009:RPP} Among the latter, FeSe has gained attention recently, in part because of the discovery of a superconducting phase above 100 K\cite{Ge:2015} for monolayers grown on appropriate substrates. Bulk FeSe exhibits a superconducting transition temperature $T_{c}$ of $9$\,K, which rises dramatically under pressure\cite{Yoshikazu:2008,Medvedev:2009}; in contrast, a single-layer FeSe film deposited on $\mathrm{SrTiO_{3}}$ substrate exhibits a $T_{c}$ increased by an order of magnitude\cite{He:2013,Tan:2013,JJ:2014,Slavko:2016}.

Like other iron chalcogenides, FeSe consists of alternating iron and chalcogenide planes, with van der Waals bonds holding together quasi-2D layers in the bulk\cite{Hsu14262,Si:2016}. When cooled across a characteristic temperature $T_{\rm S}\sim90$\,K, FeSe undergoes a nematic transition that breaks $C_4$ crystal rotational symmetry in the iron-plane with a tetragonal to orthorhombic structural transition\cite{PhysRevLett.103.057002, OriginNematicPnictide}. While the iron pnictides display a collinear striped spin-density-wave (SDW) phase immediately following a similar structural transition\cite{Kim:2011,Avci:2012,Rotundu:2011}, and other iron chalcogenides possess magnetic orders\cite{Bao:2009,Li:2009}, no long-range magnetic order has been observed for FeSe\cite{MagneticGroundStateFeSe}.

Considering the critical role that spin fluctuations may play in the unconventional, iron-based superconductors\cite{Si:2016,Li:2009:RPP,Chubukov:2012}, understanding the magnetic properties of iron chalcogenides, in particular FeSe, is helpful in identifying the nature of the pairing mechanism.  To that end, experimental evidence from neutron scattering for magnetic frustration \cite{MagneticGroundStateFeSe} and competing magnetic  ordering tendencies found in mean-field theoretical solutions of spin models\cite{Glasbrenner-Mazin} paint a picture of finely balanced interactions among various magnetically ordered phases.

Experimental and theoretical evidence suggests that despite the fact that FeSe is a metal with itinerant electrons, the low energy physics in FeSe can be described well in terms of localized electrons, owing to strong electronic correlations\cite{Tamai:2010,Yamasaki:2010,KineticFrustration},
with a fluctuating magnetic moment of $\expval{m^{2}}\sim5\mu_{B}^{2}$ per Fe atom\cite{MagneticGroundStateFeSe} corresponding to $S=1$.

A mean-field phase diagram for this type of localized electron model shows four dominant magnetic phases: N\'{e}el order [$(\pi, \pi)$], a collinear striped phase [$(\pi, 0)$ or $(0,\pi)$], a staggered dimer phase [$(\pi, \pi/2)$ and equivalent], and a double stripe phase [$(\pi/2, \pi/2)$ and equivalent]\cite{Glasbrenner-Mazin}.  Previous experiments and first-principles studies have measured spin correlations consisting of multiple wavevectors, demonstrating a magnetic frustration lacking long-range order\cite{KineticFrustration, MagneticGroundStateFeSe}. This motivates the use of a spin-1 Heisenberg model with long-range spin interactions in a regime with magnetic frustration\cite{Quantum-Paramagnetism-in-FeSe, Glasbrenner-Mazin}. Two regions of the phase diagram were previously identified as appropriate for FeSe:  a parameter regime with competition between the N\'{e}el and collinear striped orders, and one between the staggered dimer and collinear striped orders.

\section{Model and methods}

Here, we study the physics of a spin-1 Heisenberg model on a two-dimensional, 16-site cluster using exact diagonalization. Through benchmarking with mean-field theory and two different experiments, our study sets the stage for investigating the nature of FeSe within the spin model.
For parameters tuned to a frustrated region among the N\'{e}el order, staggered dimer, and collinear striped phases we evaluate the temperature dependence of the dynamical spin structure factor $S(\mathbf{q},\omega)$ and the Raman scattering cross-section. Consistent with neutron scattering,\cite{MagneticGroundStateFeSe} we find intense fluctuations of the collinear stripe order at low temperatures that give way to enhanced fluctuations at the N\'{e}el order wavevector for higher temperatures. Raman scattering\cite{Baum:2019} suggests a dominant spin character for a persistent peak in the $B_{1g}$ symmetry close to 60\,meV, which softens slightly at higher temperatures.

Due to strong electron correlations and the fluctuating magnetic moment, which neutron experiments have found to correspond well with a $S=1$ system\cite{MagneticGroundStateFeSe}, the spin-1 $J_{1}$-$J_{2}$-$J_{3}$-$K$ Heisenberg model, and similar variants, have been used to study the magnetic properties of FeSe\cite{Glasbrenner-Mazin, Quantum-Paramagnetism-in-FeSe}. The Hamiltonian can be written as
\begin{equation} \label{eq:Hamiltonian}
\begin{split}
  \mathcal{H}=\sum_{\langle i,j\rangle}\Big[J_{1}\,\mathbf{S}_{i} \cdot \mathbf{S}_{j} + K(\mathbf{S}_{i} \cdot \mathbf{S}_{j})^{2}\Big] \qquad\\ + \sum_{\langle\!\langle i,j\rangle\!\rangle} J_{2}\,\mathbf{S}_{i} \cdot \mathbf{S}_{j} + \sum_{\langle\!\langle\!\langle i,j\rangle\!\rangle\!\rangle} J_{3}\,\mathbf{S}_{i} \cdot \mathbf{S}_{j},
\end{split}
\end{equation}
where $\mathbf{S}_{i} = (S^x_i, S^y_i, S^z_i)$ is a spin operator at site $i$, $J_{\alpha} (\alpha=1,2,3)$ are the nearest, next-nearest, and next-next-nearest neighbor exchange interactions, and $K$ is the nearest-neighbor biquadratic interaction. The nearest neighbor exchange term $J_1$ favors a N\'{e}el state, while the longer-range exchange terms ($J_2$ and $J_3$) frustrate it. A large $J_2$ or $J_3$ can overwhelm $J_1$ and drives the staggered dimer or double stripe phase\cite{Glasbrenner-Mazin}.
In addition, the biquadratic term $K$ modulates fluctuations depending on the sign: a negative $K$ suppresses quantum fluctuations towards an Ising-like model\cite{Glasbrenner-Mazin}, while a positive $K$ enhances quantum fluctuations\cite{Niesen:2017} and has been found to favor a semi-ordered semi-classical ground state, containing some correlations between neighboring sites in an otherwise disordered system \cite{PAPANICOLAOU1988367}. In this work, we adopt a small positive $K$ to enhance quantum fluctuations.

We study the model on a 4$\times$4 cluster with periodic boundary conditions. This 16-site system provides access to all the relevant momenta, while remaining computationally tractable for the temperature range of interest.  While determining the ground state for such a problem is not computationally challenging, a study of the temperature dependence requires an accurate evaluation of the excited states to cover an energy spectrum in excess of the thermal energy scale set by the temperature $T$.  We adopt the parallel Arnoldi method\cite{Sorensen:1998} to determine the eigenstates and energies, and use the continued fraction expansion\cite{RevModPhys.66.763} to calculate the finite-temperature dynamical structure factor and Raman response function.

\begin{figure}[!t]
  \centering
      \includegraphics[width=\columnwidth]{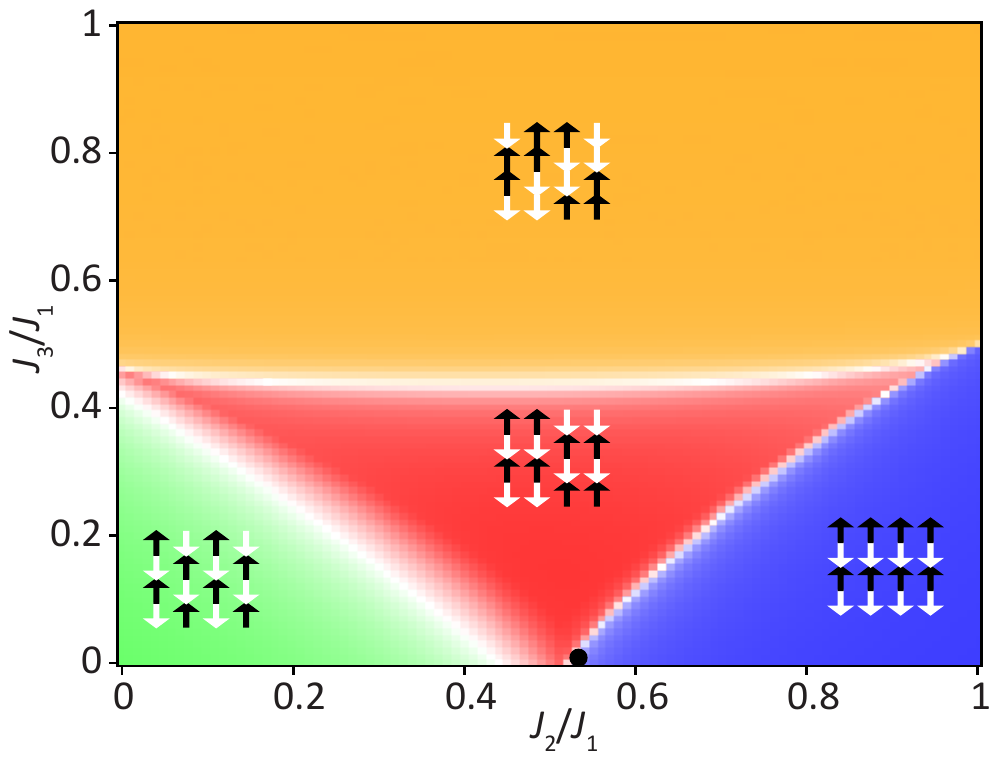}
  \caption{Zero temperature ``phase diagram'' for the $J_{1}$-$J_{2}$-$J_{3}$-$K$ spin-1 Heisenberg model with $K=0.1\,J_{1}$. The four different colors represent regions in parameter space dominated by fluctuations of the spin arrangement depicted in each cartoon, where green is N\'{e}el, red is staggered dimer, blue is collinear stripe, and orange is double stripe. The color intensity denotes the relative strength $I$, as defined in Eq.~(\ref{eq:strength}). The black circle ($J_{2}=0.528\,J_{1}$, $J_{3}=0$) denotes parameters for which we calculate the dynamical spin structure factor and Raman susceptibility as a function of temperature.}
  \label{fig:PhaseDiagram}
\end{figure}

A crucial task of this work is determining a physical set of model parameters, within the $J_{1}$-$J_{2}$-$J_{3}$-$K$ Heisenberg model, that accounts for the low-energy properties of FeSe.
To examine its dominant magnetic fluctuations and ordering instability as a function of these parameters, we first evaluate the static spin correlation function
\begin{equation} \label{eq:StaticStructureFactor}
  S(\mathbf{q}) = \frac{1}{N} \sum_{l} e^{i\mathbf{q} \cdot \mathbf{r}_{l}} \sum_{i}  \expval{\mathbf{S}_{\mathbf{r}_{i}+\mathbf{r}_{l}} \cdot \mathbf{S}_{\mathbf{r}_{i}}},
\end{equation}
where, $\mathbf{r}_{l}$ represents the coordinate of site $l$ on the cluster and the expectation value is taken with respect to the ground state at zero temperature.  To fairly parameterize the relative strength of fluctuations with different characteristic wavevectors, we normalize the relative intensity of the dominant and largest subdominant correlation functions. Thus, the relative strength of fluctuations is projected onto the range [0,1) by
\begin{equation} \label{eq:strength}
  I = 1 - \frac {d_{\mathbf{q}_{\mathrm{sub}}}S(\mathbf{q}_{\mathrm{sub}})}{d_{\mathbf{q}_{\mathrm{dom}}} S(\mathbf{q}_{\mathrm{dom}})}
\end{equation}
where $d_{\mathbf{q}}$ is the geometric degeneracy for each equivalent momentum point on the 4$\times$4 cluster and $\mathbf{q}_{\mathrm{dom/sub}}$ denote the value of $\mathbf{q}$ for which $d_{\mathbf{q}}S(\mathbf{q})$ is largest (dominant) / second largest (subdominant). Figure~\ref{fig:PhaseDiagram} shows the resulting ``phase diagram'', which displays the order with the dominant correlation not the true long-range order of the system, obtained in this manner for the $J_{1}$-$J_{2}$-$J_{3}$-$K$ model. Clearly, in contrast to the N\'{e}el order state (green) in the canonical Heisenberg model, the next-nearest neighbor exchange $J_2$ favors a collinear striped state (blue) while the longer-range $J_3$ stabilizes a double stripe state (orange) above some critical couplings. In the middle of these three states, the combined impact of exchange interactions induces a staggered dimer region (red). Near the boundaries large fluctuations due to frustration suppress the states (white regions).

Given the various instabilities of the $J_{1}$-$J_{2}$-$J_{3}$-$K$ model, can one find a parameter regime appropriate for FeSe? Previously, Fa~Wang \textit{et al.}~adopted a $J_{1}$-$J_{2}$ model near the quantum paramagnetic phase around $J_{2}\! \sim\! 0.5\,J_{1}$\cite{Quantum-Paramagnetism-in-FeSe} and Qisi~Wang \textit{et al.}~suggested a point in the staggered dimer region near the boundary with collinear striped order (for negative $K$)\cite{MagneticGroundStateFeSe}. Both involve competition between collinear striped order and some other state. This would be consistent with recent neutron scattering data, showing both collinear striped and, slightly weaker, N\'{e}el order fluctuations at low temperatures, with spectral weight transfer between them upon changing the temperature.  Such an experimental observation suggests that the low-energy magnetic properties of FeSe can be described by a parameter set inside the collinear stripe region close to its boundary.

\section{Spectral results}

In addition to neutron scattering, the Raman response provides another clue about a proper para\-meter regime for FeSe as it captures the two-magnon excitations. As a collective mode, the two-magnon excitations depend sensitively on the form and strength of magnetic interactions\cite{Fleury:1968}.  At low energy, the experimental Raman response in $B_{1g}$ symmetry consists of two dominant contributions which can be separated in the temperature range around $T_{\rm S}$ [\onlinecite{Baum:2019}]. The peak in the range below 200\,cm$^{-1}$ was interpreted previously in terms of charge nematic fluctuations \cite{Massat:2016}. We return to this point briefly later.  Here we focus on the broad peak centered at 500\,cm$^{-1}$ which softens slightly and loses weight with increasing temperature\cite{Baum:2019}. We argue that this part of the Raman response originates from spin excitations and will elaborate now on the theoretical details.

In the Fleury-Loudon formalism,\cite{Fleury:1968} the Raman scattering operator is written as $\hat{O}=\sum_{i,j} J_{ij} (\hat{\mathbf{e}}_{\mathrm{in}} \cdot \hat{\mathbf{d}}_{ij}) (\hat{\mathbf{e}}_{\mathrm{out}} \cdot \hat{\mathbf{d}}_{ij})\,\mathbf{S}_{i} \cdot \mathbf{S}_{j}$\cite{CCCRaman}, where $J_{ij}$ are exchange coupling strengths in the spin Hamiltonian, $\hat{\mathbf{d}}_{ij}$ represent unit vectors connecting sites $i$ and $j$,  and $\hat{\mathbf{e}}_{\mathrm{in/out}}$ are the polarization vectors for the incoming/outgoing photons, respectively. The light polarizations that encode the Raman symmetry channels are
\begin{equation}
\left\{\begin{array}{ll}
\hat{\mathbf{e}}_{\mathrm{in}}=\frac{1}{\sqrt{2}}(\hat{x}+\hat{y}), \hat{\mathbf{e}}_{\mathrm{out}}=\frac{1}{\sqrt{2}}(\hat{x}+\hat{y}) &\textrm{ for } A'_{1g},\\
\hat{\mathbf{e}}_{\mathrm{in}}=\hat{x}, \hat{\mathbf{e}}_{\mathrm{out}}=\hat{y}& \textrm{ for } B_{2g},\\
\hat{\mathbf{e}}_{\mathrm{in}}=\frac{1}{\sqrt{2}}(\hat{x}+\hat{y}), \hat{\mathbf{e}}_{\mathrm{out}}=\frac{1}{\sqrt{2}}(\hat{x}-\hat{y}) &\textrm{ for } B_{1g},
\end{array}\right.
\end{equation}
where $A'_{1g}=A_{1g} \oplus B_{2g}$. In this work, we mainly focus on the $B_{1g}$ channel as it directly reveals the two-magnon excitation, while the $A_{1g}$ and $B_{2g}$ spectra serve as additional experimental comparison.

\begin{figure}[!th]
  \centering
      \includegraphics[width=\columnwidth]{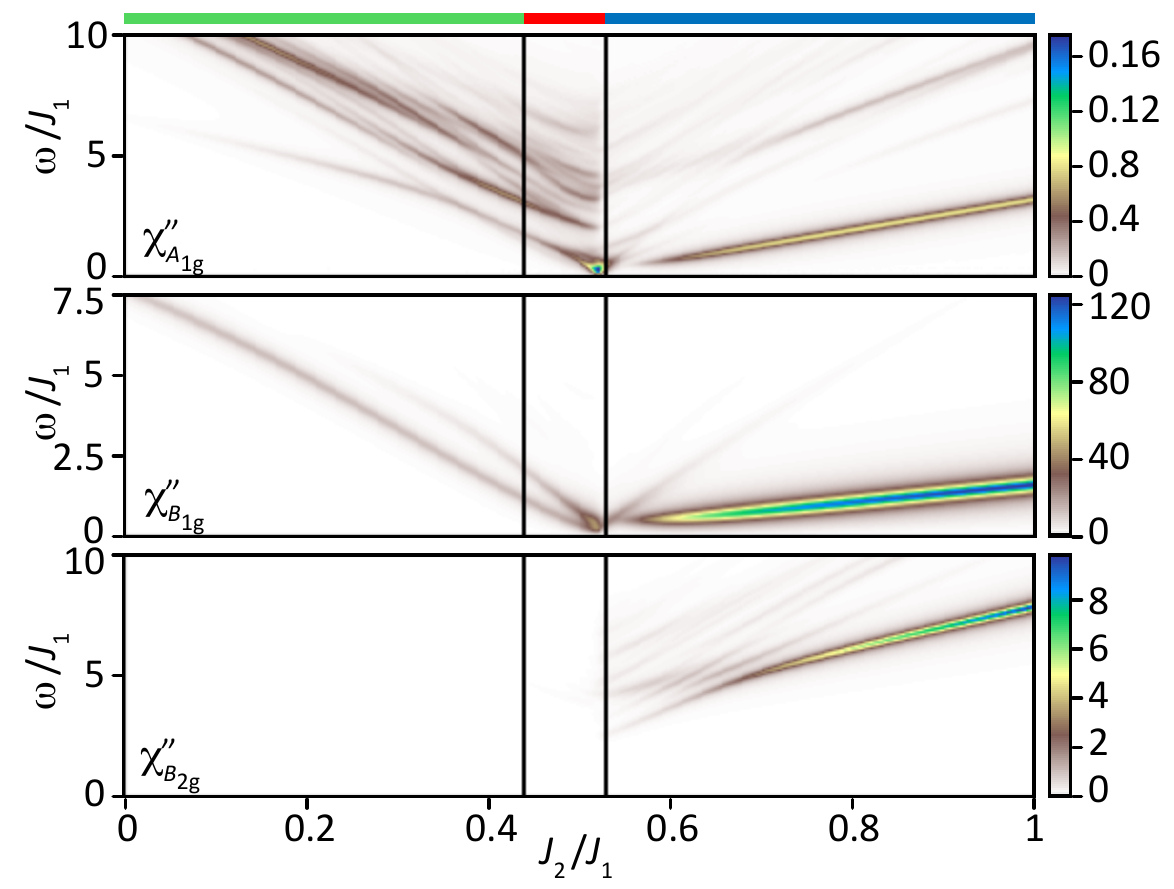}
  \caption{Raman susceptibility $\chi''(\omega)$ at zero temperature as a function of $J_{2}$, with $J_{3}=0$ and $K=0.1J_{1}$ for $A_{1g}$ (top), $B_{1g}$ (middle), and $B_{2g}$ (bottom) symmetries. The two vertical black boundary lines and the top color bar sketch regions with distinct dominant correlations [N\'{e}el order (left), staggered dimer (middle), and collinear stripe (right), as in Fig.~\ref{fig:PhaseDiagram}]. }
  \label{fig:RamanJ2}
\end{figure}

Using the Raman scattering operator $\hat{O}_{\alpha}$, we evaluate the temperature-dependent Raman response in different symmetry channels as
\begin{equation} \label{eq:R(w)}
R_{\alpha}(\omega)\! =
 - \!\sum \limits_{n}\! \frac{e^{-\!\beta\! E_{n}}}{\pi Z}\!  \operatorname{Im}\!\matrixel{\psi_{n}}{\hat{O}_\alpha^\dagger W^{-1} \hat{O}_{\alpha}}{\psi_{n}},
\end{equation}
where $\alpha$ denotes a particular symmetry channel, $Z$ is the partition function, $W\,=\,\omega \!+\! E_{n}\! +\! i\epsilon\! -\! \mathcal{H}$, $\ket{\psi_{n}}$ and $E_{n}$ are the $n$-th eigenstate and energy, with the sum taken over all eigenstates of the Hamiltonian in Fock space\cite{RevModPhys.66.763}. We use $\epsilon=0.15J_{1}$ in the continued fraction step. To remove the elastic peak, it is convenient to calculate the Raman susceptibility $\chi^{\prime\prime}_{\alpha}(\omega)\! =\! R_{\alpha}(\omega)\! -\! R_{\alpha}(-\omega)$.  Due to the computational challenges, we truncate the summation at an energy $E_{0}+2J_{1}$, while providing sufficient states for evaluating the temperature-dependence of spectra up to $T=0.25J_1$ (all states contributing a weight $e^{-\beta E_n}>e^{-5}$).

As shown in Fig.~\ref{fig:RamanJ2}, the Raman susceptibility at zero temperature changes dramatically with $J_{2}$. In $B_{1g}$ symmetry (middle panel)  the two-magnon excitation starts around an energy of $7.5J_{1}$ for $J_{2}=0$, then softens uniformly approaching the boundary between the staggered dimer and collinear striped phases. The energy for this two-magnon excitation can be estimated by counting the number of interactions that change sign with a double spin flip. Across the transition from N\'{e}el to staggered dimer, the transition is gradual as demonstrated by the wide white region in Fig.~\ref{fig:PhaseDiagram}. In contrast, the transition across staggered dimer and collinear stripe order is more abrupt and leads to discontinuous changes in the Raman spectra. Taking a value of $J_{1}=123.1\,$meV from first principles calculations\cite{Glasbrenner-Mazin}, it becomes clear that  consistency between the experimental position of the peak at roughly $500\,\textrm{cm}^{-1}$ and the theoretical two-magnon energy can only be obtained for $J_2/J_1 \sim 0.5$, near the boundary between the staggered dimer and collinear striped phases. This parameter range is also consistent with the general notion of highly frustrated magnetism in FeSe. We identify the best agreement in this region with $J_{2}=0.528J_{1}$, $J_{3}=0$, and $K=0.1J_{1}$ (the black dot shown in Fig.~\ref{fig:PhaseDiagram}). The  significance of these parameters is presumably the positive biquadratic coupling $K$ and the $J_{2}$ value that puts the system very close to the phase boundary in the immediate vicinity of the collinear stripe region. The exact numerical values of the parameters that describe FeSe are expected to change slightly in other finite size clusters and in the thermodynamic limit. As we show next, the finite temperature Raman and neutron scattering experiments compare favorably with simulations for these parameters.

\begin{figure}[!t]
  \centering
      \includegraphics[width=\columnwidth]{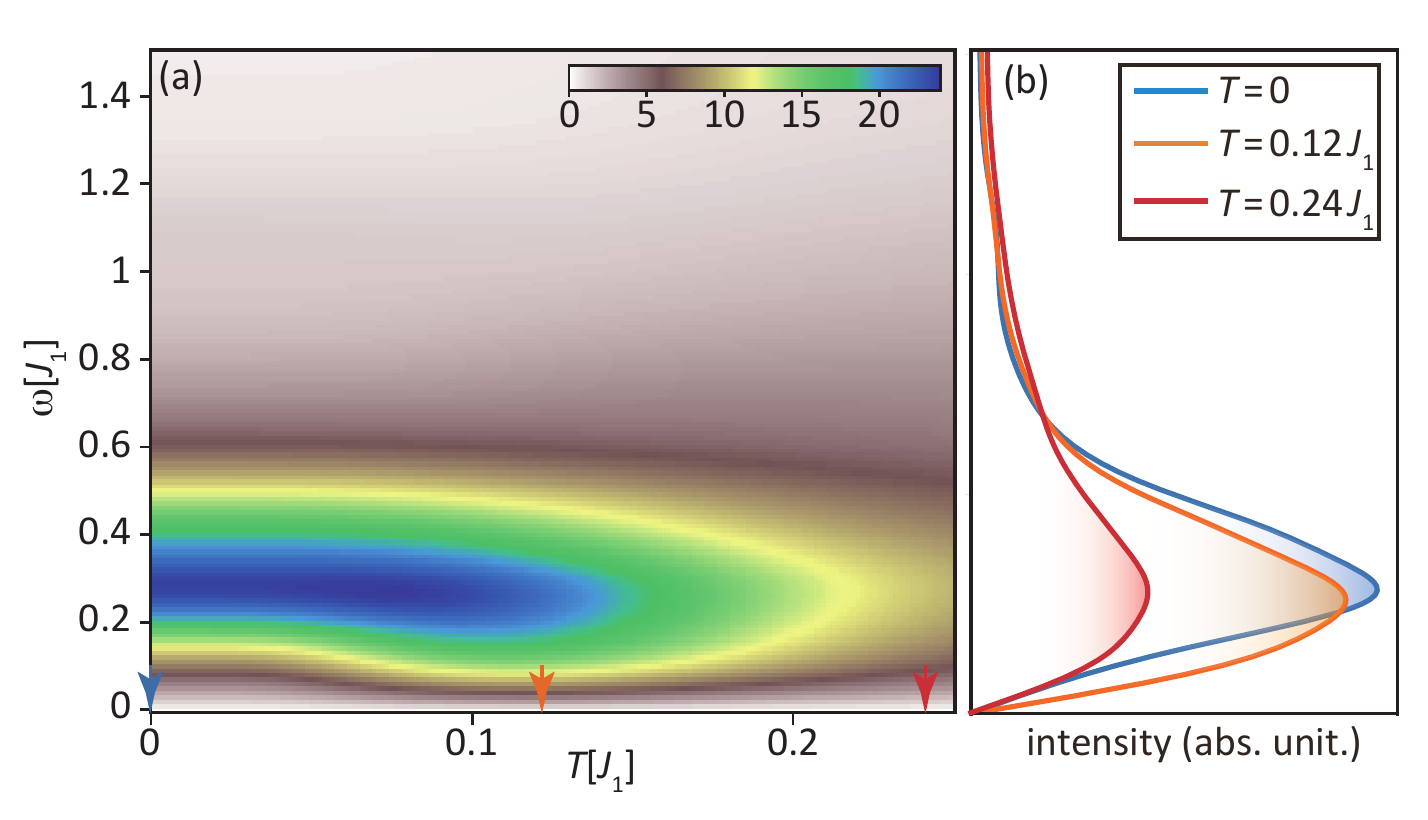}
  \caption{(a) The imaginary part of the Raman susceptibility for $B_{1g}$ symmetry as a function of temperature. (b) Cuts of Raman spectra at $T=0$ (blue), $0.12J_1$ (orange), and $0.24J_1$ (red), indicated by the arrows.
  }   \label{fig:Raman}
\end{figure}

Figure~\ref{fig:Raman} displays the temperature dependence of the $B_{1g}$ Raman susceptibility for the chosen parameters. We observe a single dominant peak. With increasing temperature, this peak softens slightly, before hardening again at higher temperature. The peak gradually loses intensity up to the highest simulated temperatures. In fact, thermal broadening occurs in all symmetries, while $B_{1g}$ remains dominant. This dominant peak that softens with increasing temperature before hardening again agrees well with experiment\cite{Baum:2019}, suggesting that a local spin model provides an adequate description of the dominant degrees of freedom in Raman scattering from FeSe in this energy range. We will see that the temperature dependence of this softening coincides with the temperature dependence of spectral weight transfer observed in $S(\mathbf{q}=(\pi,\pi),\omega)$, further reinforcing the connection between the peak in the Raman response and two magnon excitations.

The spectral range below the magnon excitations is dominated by critical fluctuations peaking at 50\,cm$^{-1}$ close to $T_{\rm S}$ [\onlinecite{Massat:2016,Baum:2019}]. The origin of these fluctuations is not obvious. While Massat \textit{et al.} argue for orbital (charge) fluctuations\cite{Massat:2016} critical spin fluctuations cannot \textit{a priori} be excluded in the ubiquitous presence of magnetism. Yet, our simulations in the spin channel do not support this interpretation. However, critical fluctuations cannot be captured by a simulation on a $4\times4$ cluster since they are characterized by a diverging correlation length for $T\to T_{\rm S}$. The shoulder on the low-energy side of the magnon excitation may be a remainder of the fluctuations but further studies are necessary.

\begin{figure}[!ht]
  \centering
      \includegraphics[width=1.0\columnwidth]{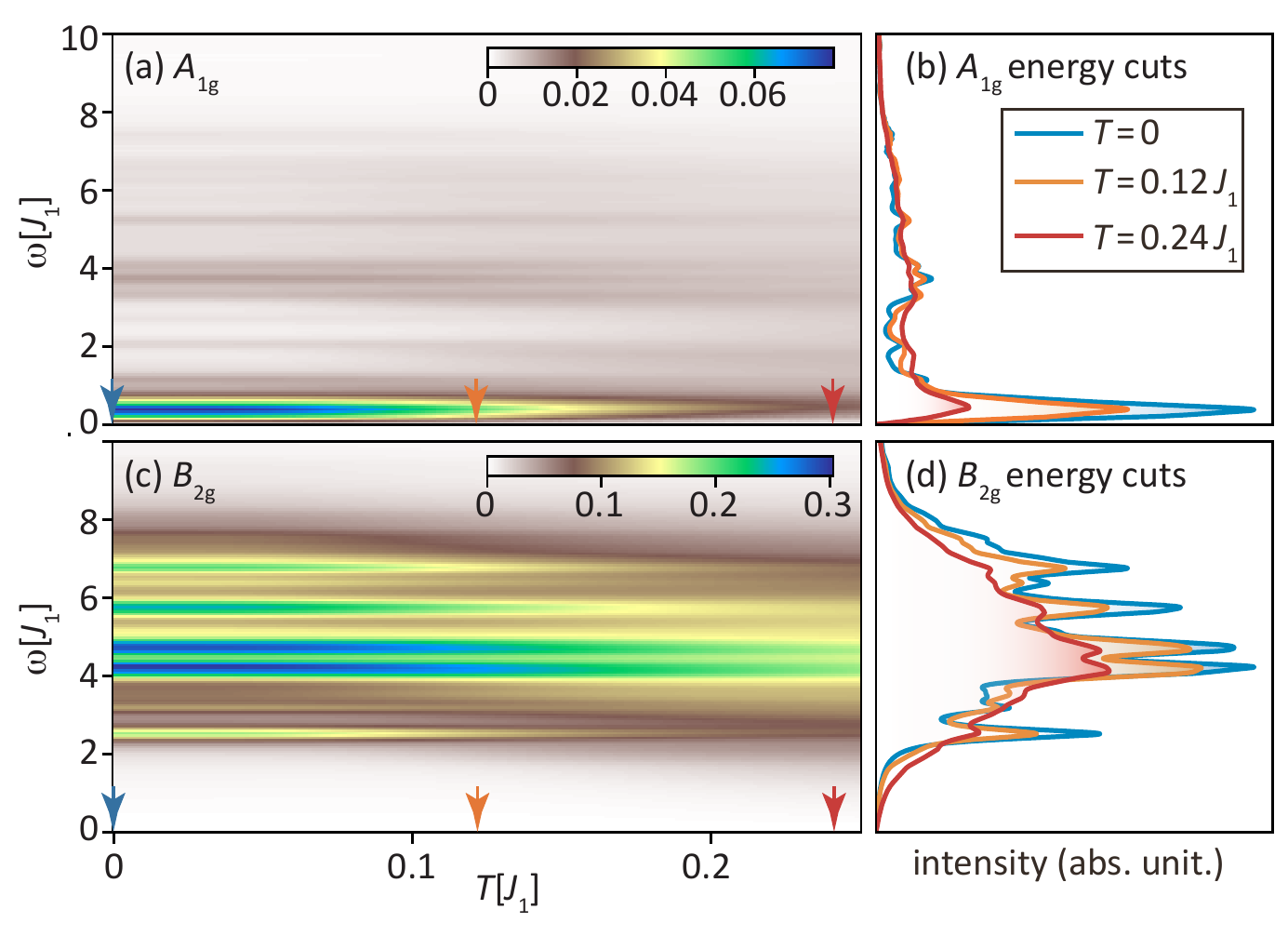}
  \caption{The imaginary part of the Raman susceptibility for (a) $A_{1g}$ and (c) $B_{2g}$ symmetries at temperatures as indicated. (b,d) Cuts corresponding to $A_{1g}$ and $B_{2g}$ Raman spectra at three temperatures $T=0$, 0.12$J_1$ and 0.24$J_1$.
  }
  \label{fig:Raman2}
  \end{figure}

In $A_{1g}$ symmetry [Fig.~\ref{fig:Raman2}(a)], there is a single peak at slightly higher energy than the one found in the $B_{1g}$ channel. This peak decreases in intensity and hardens slightly with increasing temperature. In $B_{2g}$ symmetry, shown in Fig.~\ref{fig:Raman2}(b), we see several peaks spread out over the energy range of $2J_{1}$ to $8J_{1}$.  The general trends for each of these symmetries, and in particular the dominant peak in $B_{1g}$ symmetry, correspond well with recent Raman scattering data.\cite{Baum:2019}

\begin{figure}[!ht]
  \centering
      \includegraphics[width=1.0\columnwidth]{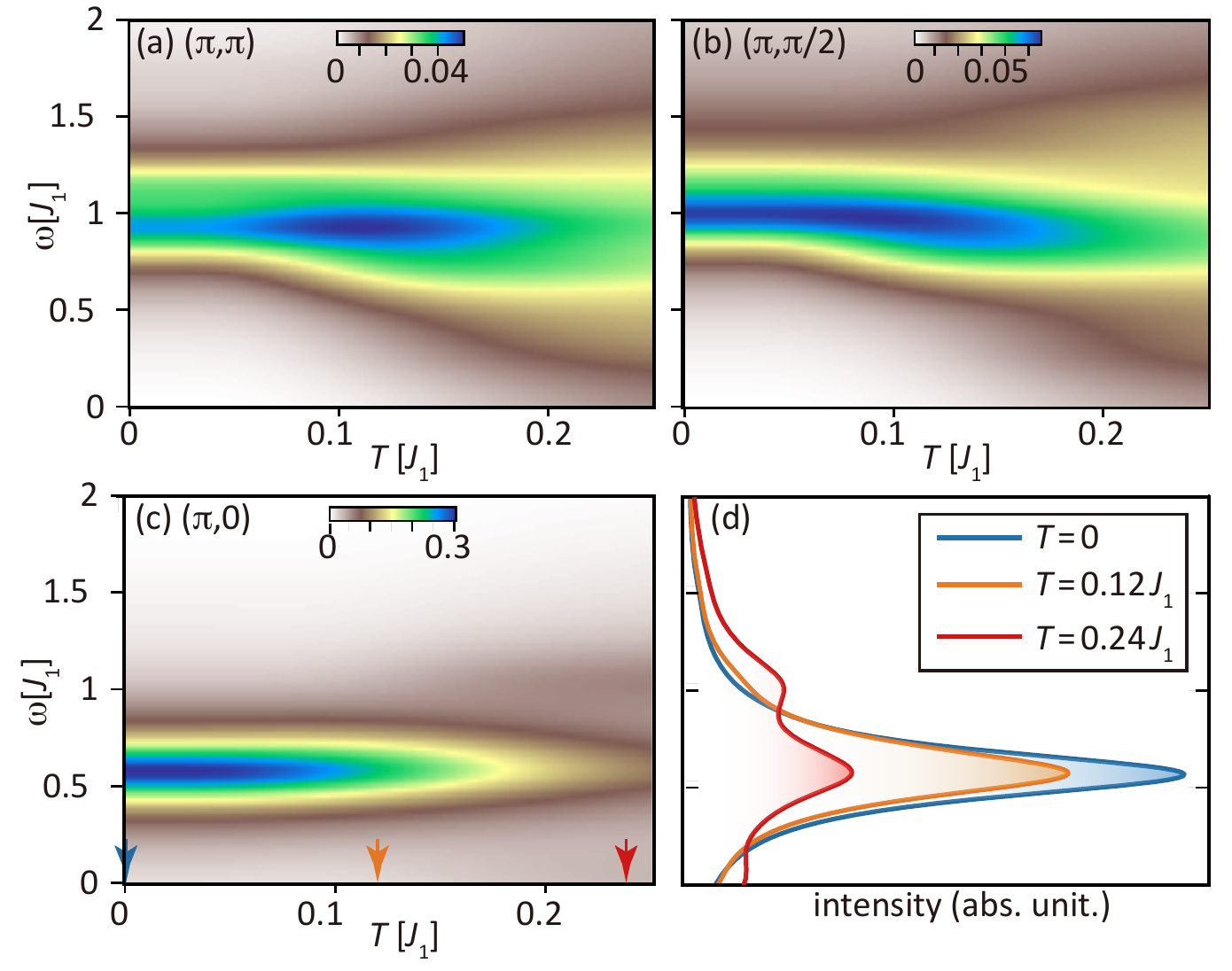}
  \caption{(a-c) $S(\mathbf{q}, \omega)$ as a function of temperature for $\mathbf{q}=(\pi, \pi)$ (N\'{e}el), $\mathbf{q}=(\pi, \frac{\pi}{2})$ (staggered dimer) and $\mathbf{q}=(\pi, 0)$ (collinear stripe), respectively.  More detailed temperature dependence can be seen for three temperature cuts of $(\pi, 0)$ in (d). As temperature increases, spectral weight shifts to lower energy at $(\pi, \pi)$ and $(\pi, \frac{\pi}{2})$, and to higher energy at $(\pi, 0)$.
  }
  \label{fig:SqwStacked}
\end{figure}

While we have seen that Raman scattering provides some information about magnetic excitations in the model, much more detailed information comes from the dynamical spin structure factor
\begin{equation} \label{eq:S(q,w)}
S(\mathbf{q}, \omega)\! =\! -\! \sum \limits_{n}\! \frac{e^{-\!\beta\! E_{n}}}{\pi Z}\!  \operatorname{Im}\matrixel{\psi_{n}}{S_{\mathbf{-q}}^{z} W^{-1}  S_{\mathbf{q}}^{z}}{\psi_{n}},
\end{equation}
where $S_{\mathbf{q}}^{z} = \frac{1}{\sqrt{N}}\sum_{l}e^{i\mathbf{q} \cdot \mathbf{r}_{l}} S_{l}^{z}$. Figure~\ref{fig:SqwStacked} shows $S(\mathbf{q}, \omega)$  as a function of temperature for $\mathbf{q}=(\pi,0)$, $(\pi,\pi)$, and  $(\pi,\frac{\pi}{2})$. At $T=0$, the lowest-energy spin excitation occurs at $(\pi,0)$, with significant fluctuations at slightly higher energy in $(\pi,\frac{\pi}{2})$ and $(\pi,\pi)$, indicative of a frustrated magnetic system.
With increasing temperature, the spin excitation at $(\pi,0)$ hardens slightly and loses intensity, while it softens substantially at $(\pi, \pi)$ and $(\pi, \frac{\pi}{2})$. This temperature dependence is reminiscent of the neutron scattering data\cite{MagneticGroundStateFeSe}, and the enhanced competition is consistent with the evolution of the $B_{1g}$ Raman response in Fig.~\ref{fig:Raman}, further highlighting the role of magnetic frustration in FeSe.

\section{Discussion}

\begin{figure}[!t]
  \centering
      \includegraphics[width=9cm]{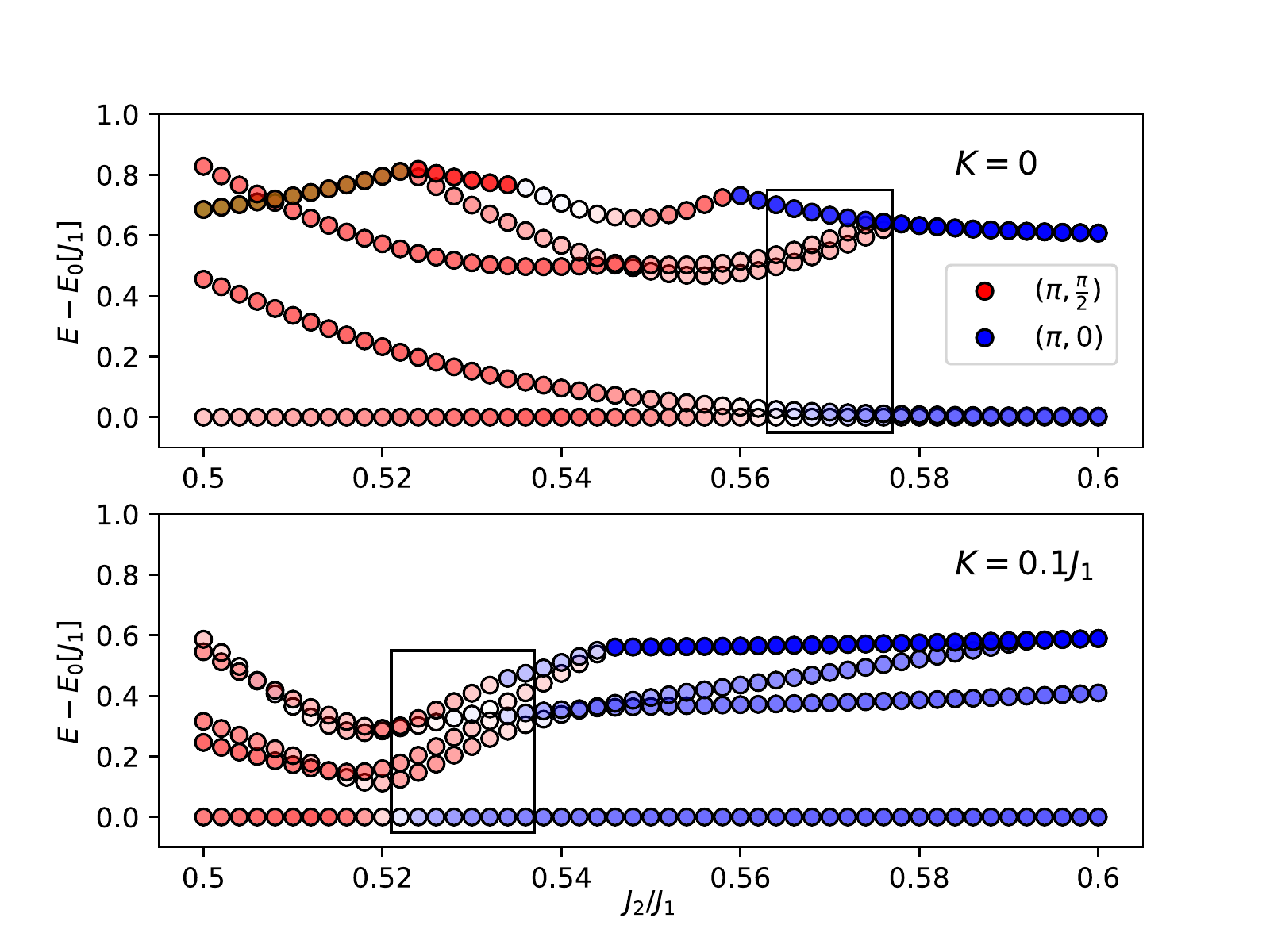}
  \caption{Energy and magnetic fluctuations associated with the five lowest energy excited states for $K=0$ (top) and $0.1J_{1}$ (bottom), as a function of $J_{2}$ for $J_{3}=0$. The black boxes enclose a range of $J_{2}$ where the ground state and possibly a nearly degenerate state of collinear striped order are followed by states characterized by a dominant staggered dimer phase. The color coding of each circle follows the same convention as Fig.~\ref{fig:PhaseDiagram}. }
  \label{fig:Evals}
\end{figure}

Interestingly, only a small region of parameter space with $K \sim 0.1J_{1}$ displays a temperature dependence consistent with the $B_{1g}$ Raman and the spin response, at least in this 16-site cluster calculation (Raman response functions for other parameters shown in the supplementary material). The origin of this softening and its sensitivity to $K$ is difficult to assess in a simple spin-wave picture due to the many-body nature of this biquadratic term. Fortunately, with the full wavefunctions obtained by exact diagonalization, we can study the magnetic fluctuations and competition directly through eigenstates of the Hamiltonian.
Figure~\ref{fig:Evals} shows detailed information about the five lowest eigenstates as a function of $J_{2}$ for two different values of $K$. The color of each point represents the dominant magnetic character of the eigenstate, following the same convention as Fig.~\ref{fig:PhaseDiagram}. Crossing the boundary to the collinear striped phase, there is a small region (highlighted  by  the black boxes) where the low-lying excited states possess a staggered dimer or mixed character. In this region, while both values of $K$ result in similar ground states and zero-temperature Raman and neutron scattering spectra, only $K=0.1J_{1}$ provides the ingredients for a temperature dependence consistent with experiments, because of its much smaller excitation gap and larger density of excited states. These states are responsible for the softening of the $B_{1g}$ peak, as well as the energy shift and weight transfer of the dynamical spin structure factor.

\section{Conclusion}

In summary, we present a systematic exact-diagonalization study of the magnetic fluctuations and spectra in a local spin $J_{1}$-$J_{2}$-$J_{3}$-$K$ model. This model displays a rich phase diagram influenced by magnetic frustration. A comparison of the dynamical spin and Raman response to experimental results underscores that this model provides a consistent description of the magnetic properties of FeSe, lying at the boundary between the collinear stripe, N\'{e}el order, and staggered dimer phases. Through a detailed analysis of the eigenstates, we attribute the temperature evolution of the spectra to the competition between various finely balanced magnetic ground and excited states, and hence explain the crucial role of the biquadratic coupling. Our results suggest that magnetic frustration plays a dominant role in the low-energy physics of FeSe, which may additionally support the intrinsic connection between spin fluctuations and unconventional superconductivity. We find that local spins give an adequate description of these magnetic properties.

\section*{ACKNOWLEDGEMENTS}
Supported by the U.S. Department of Energy (DOE), Office of Basic Energy Sciences, Division of Materials Sciences and Engineering, under contract DE-AC02-76SF00515. Computational work was performed using the resources of the National Energy Research Scientific Computing Center, under contract DE-AC02-05CH11231.
A.B. and R.H. acknowledge support by the German Research Foundation (DFG) via the Transregional Collaborative Research Center TRR80 and Research Grant no. HA2071/12-1. The collaboration with Stanford University was supported by the Bavaria California Technology Center BaCaTeC (grant-no. 21 [2016-2]).

\appendix
\setcounter{figure}{0}
\renewcommand\thefigure{A\arabic{figure}}
\section{Line shape of the $B_{1g}$ Raman spectra of ${\rm FeSe}$}
\label{sec:FeSe}
\begin{figure*}[!ht]
  \centering
  \includegraphics[width=15cm]{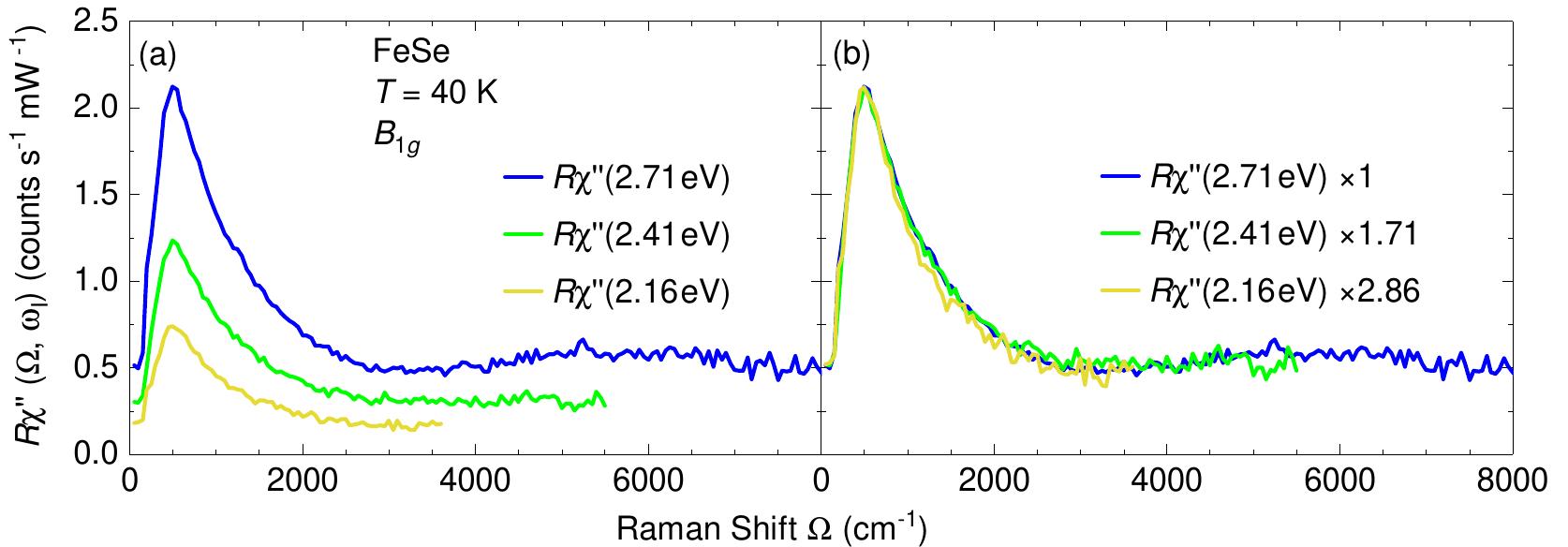}
  \caption{$B_{1g}$ Raman response $R\chi^{\prime\prime}(\Omega,\hbar\omega_I)$ of FeSe for three photon energies $\hbar\omega_I$ as indicated. All spectra are corrected for the instrumental response. (a) Raw data. At low energies the magnetic response is superimposed on the particle-hole continuum (see also supplemental section 4 of Ref.~\onlinecite{Baum:2019}). At high energies, $\Omega>2\,000$\,cm$^{-1}$, there is a substantial contribution from luminescence. (b) After appropriately multiplying the spectra measured with green and yellow photons with constant factors all spectra collapse on top of each other.
  }
  \label{Afig:FeSe-resonance}
\end{figure*}
The interaction of photons and spin excitations may be described by the Fleury-Loudon Hamiltonian $\mathcal{O}$. \cite{Fleury:1968} The Raman response is then determined as described in Eq.~(5) of the main text. The Fleury-Loudon formalism is justified only in the non-resonant case. If the intermediate electronic states are eigenstates of the band structure not only the intensity but also the line shape may depend on the energy of the photons \cite{Shastry:1990,Chubukov:1995a,Chubukov:1995b}.

For justifying the applicability of the formalism we measured the $B_{1g}$ Raman response of FeSe for three different excitation energies $\hbar\omega_I$, 2.16\,eV (575\,nm), 2.41\,eV (514\,nm), and 2.71\,eV (458\,nm). [$\lambda_{I}\,{\rm (nm)}=1240/\hbar\omega_I\,{\rm (eV)}$] The results are shown in Fig.~\ref{Afig:FeSe-resonance}. Panel (a) displays the raw data at 40\,K after correcting for the spectral response of the system. All spectra peak at approximately 530\,cm$^{-1}$, and the maxima are asymmetric having a much slower decay on the high-energy side than at low energies. The overall intensity increases by a factor of almost three if $\hbar\omega_I$ increases from 2.16 to 2.71\,eV. Panel (b) shows that the line shape is independent of the excitation energy. All spectra collapse on top of each other when multiplied appropriately. Thus the line shape does not depend on $\hbar\omega_I$. Consequently, the response derived via Eq.~(5) of the main text is qualitatively correct. The experimental and theoretical results are compared in Figs. 3 and 5 of Ref.~\onlinecite{Baum:2019}.

\setcounter{figure}{0}
\renewcommand\thefigure{B\arabic{figure}}
\begin{figure}[!ht]
  \centering
      \includegraphics[width=\columnwidth]{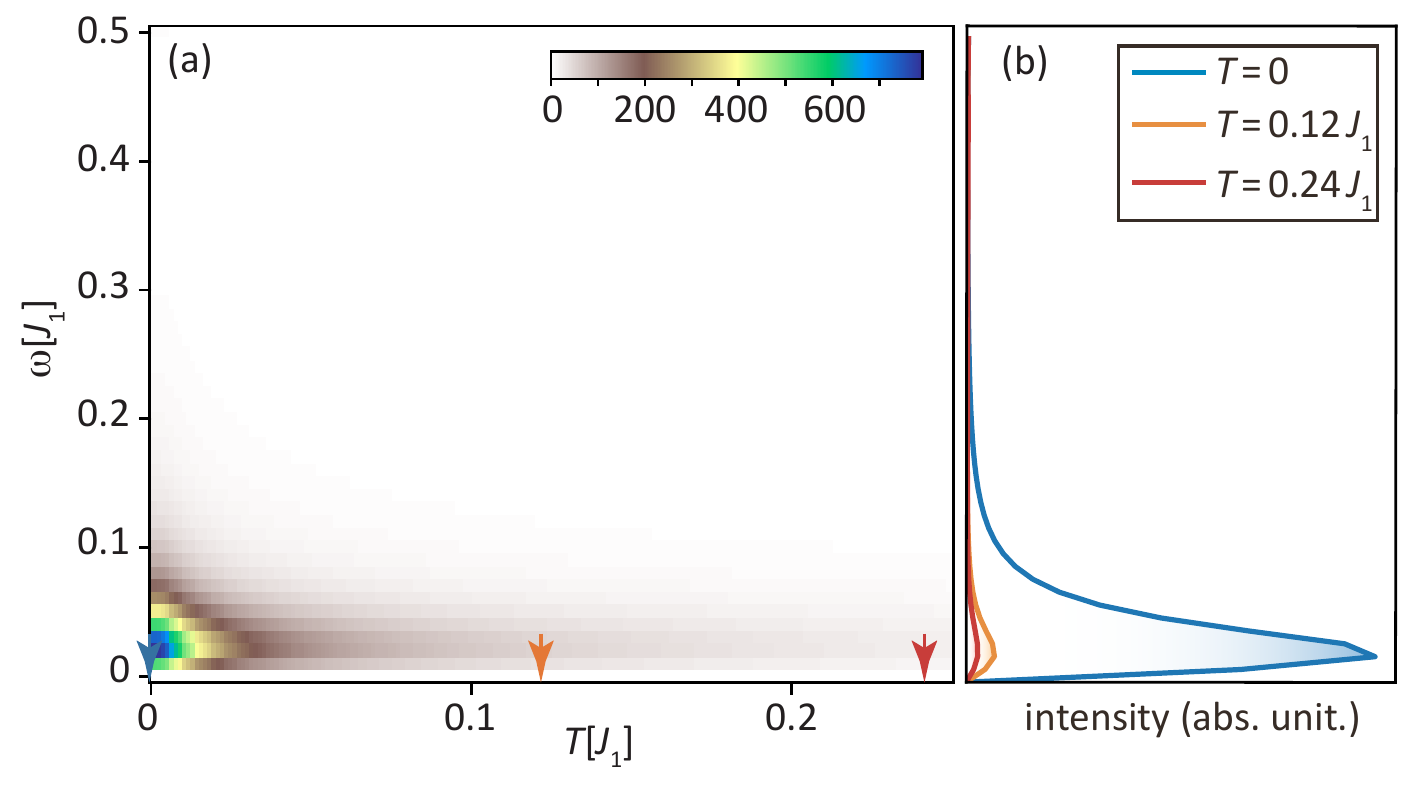}
  \caption{The imaginary part of the Raman susceptibility for $B_{1g}$ symmetry for $K=0$, $J_{2}=0.57J_{1}$, and $J_{3}=0$. These parameters are found in the collinear stripe region near the transition to staggered dimer.}
  \label{fig:K0B1g}
\end{figure}
\section{Biquadratic Coupling Dependence of Raman Spectra}
\begin{figure}[!t]
  \centering
      \includegraphics[width=0.7\columnwidth]{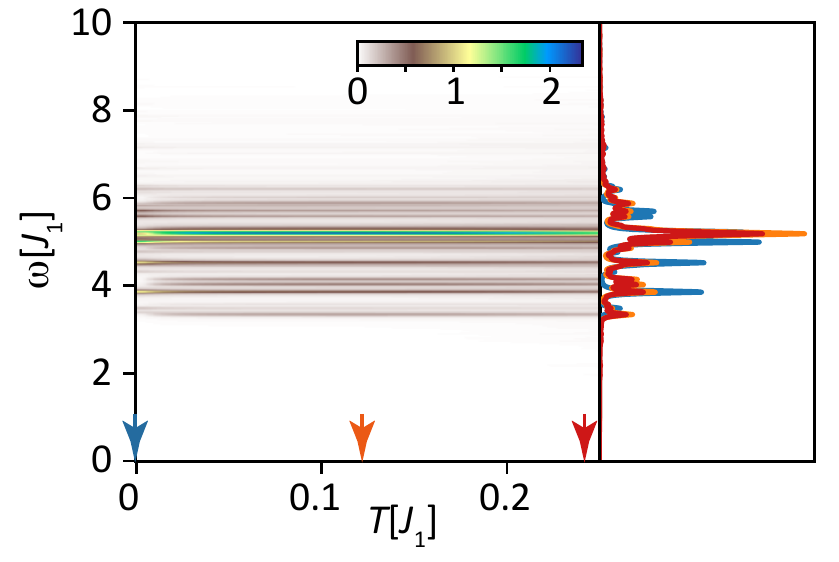}
  \caption{The imaginary part of the Raman susceptibility for $B_{2g}$ symmetry for $K=0$, $J_{2}=0.57J_{1}$, and $J_{3}=0$. These parameters are found in the collinear stripe region near the transition to staggered dimer.}
  \label{fig:K0B2g}
\end{figure}

We have found the positive biquadratic coupling to be critical to the temperature dependence of the Raman susceptibility simulations agreeing well with experiment. Here we show nonzero temperature simulations for $K=0$ and $K=0.2J_{1}$. We have picked parameters immediately inside the collinear stripe region near the transition to staggered dimer along $J_{3}=0$, similar to the point highlighted in Fig. 1 of the main text.

Fig. \ref{fig:K0B1g} shows the $B_{1g}$ Raman susceptibility for $K=0$. This was calculated in the same way as in the main text except that we used $\epsilon=0.03J_{1}$ since the energy levels are more closely packed. The spectrum still consists of a dominant low energy peak but this peak does not soften as temperature is increased, a signature of Raman scattering in FeSe in $B_{1g}$ symmetry. In addition, the maximum is at a much lower energy than for $K=0.1$. Fig. \ref{fig:K0B2g} shows the $B_{2g}$ Raman susceptibility for the same parameters. This spectrum is similar to what we see for the parameters used in the main text. We do not show the $A_{1g}$ susceptibility for these parameters since it is zero with the Raman operator we have used when $K=0$.

\begin{figure}[!t]
  \centering
      \includegraphics[width=\columnwidth]{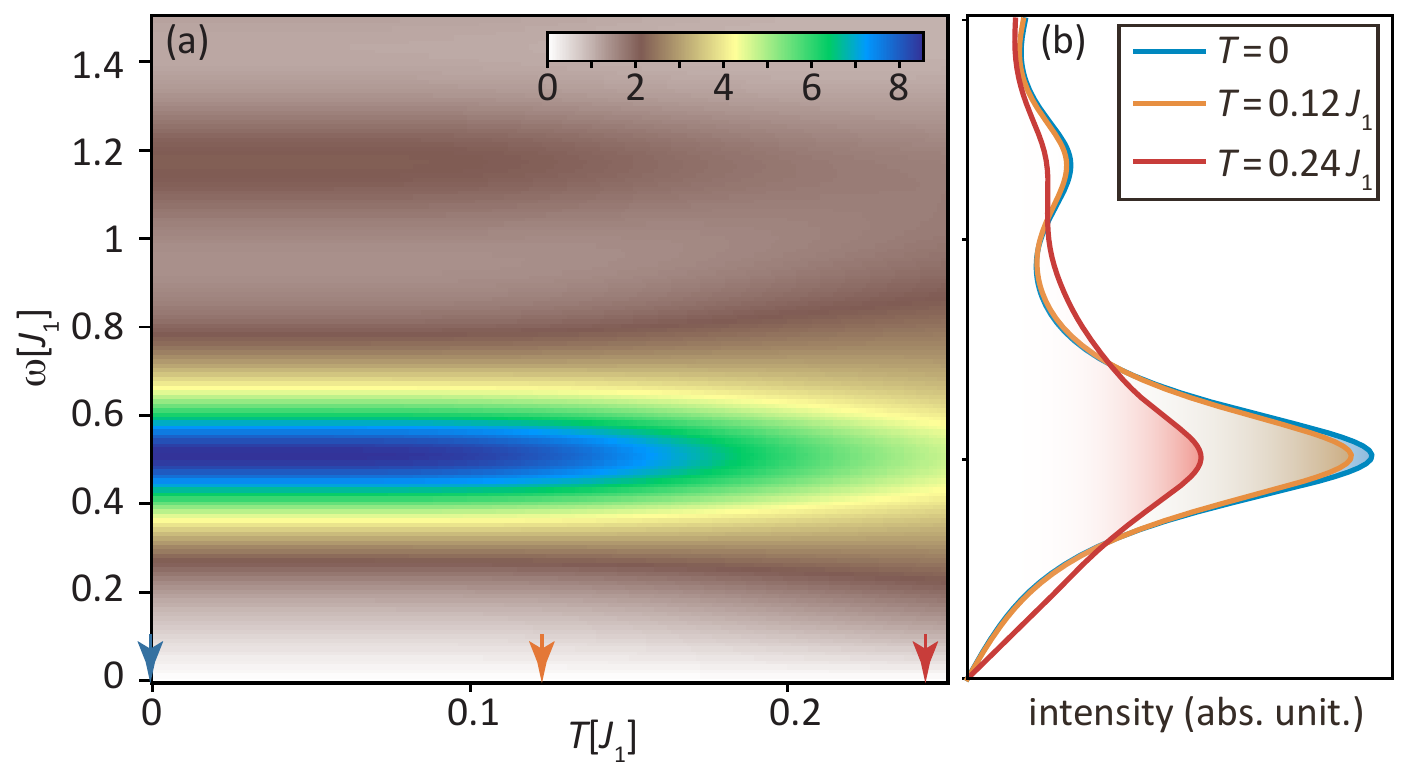}
  \caption{The imaginary part of the Raman susceptibility for $B_{1g}$  symmetry for $K=0.2J_{1}$, $J_{2}=0.47J_{1}$, and $J_{3}=0$. These parameters are found in the collinear stripe region near the transition to staggered dimer.}
  \label{fig:K200B1g}
\end{figure}

\begin{figure}[!t]
  \centering
      \includegraphics[width=\columnwidth]{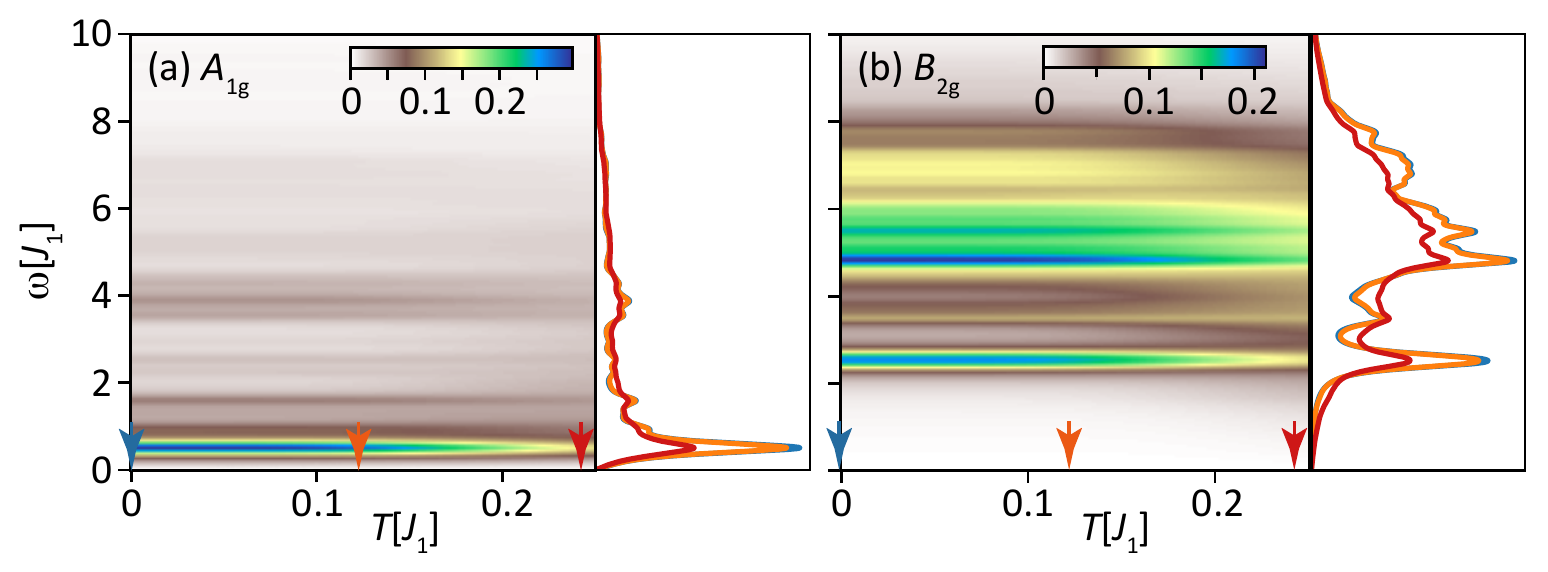}
  \caption{The imaginary part of the Raman susceptibility for $K=0.2J_{1}$, $J_{2}=0.47J_{1}$, and $J_{3}=0$ for $A_{1g}$ (a) and $B_{2g}$ (b) symmetries. These parameters are found in the collinear stripe region near the transition to staggered dimer.}
  \label{fig:K200A1gB2g}
\end{figure}

Fig. \ref{fig:K200B1g} shows the $B_{1g}$ Raman susceptibility for $K=0.2J_{1}$, again immediately inside the collinear stripe region. Here we see again a single peak that does not soften with increasing temperature. Fig \ref{fig:K200A1gB2g} shows the $A_{1g}$ and $B_{2g}$ susceptibilities. These are similar to the results shown in the main text with a single low energy peak in $A_{1g}$ symmetry and a more spread out spectrum for $B_{2g}$ symmetry. Again we see that the temperature dependence of the $B_{1g}$ Raman susceptibility is what distinguishes the biquadratic coupling parameter used in the main text from other values.

\bibliography{./Bibliography}
\clearpage
\end{document}